\documentclass[
    twocolumn,
    superscriptaddress,
    amsmath,
    amssymb,
    amstext,
    aps,
    pra,
    longbibliography,
]{revtex4-1}

\usepackage[colorlinks=true,citecolor=blue,linkcolor=blue, urlcolor=blue]{hyperref}

\usepackage{
	xcolor,
    graphicx,  
}

\usepackage{colortbl}
\colorlet{g}{green!0!}
\colorlet{y}{yellow!0!}
\colorlet{r}{red!0!}
\begin{document}

\title{Extending quantum probabilistic error cancellation by noise scaling}

\author{Andrea Mari}
\affiliation{Unitary Fund}
\author{Nathan Shammah}
\affiliation{Unitary Fund}
\author{William J. Zeng}
\affiliation{Unitary Fund}
\affiliation{Goldman, Sachs \& Co, New York, NY}


\begin{abstract}
We propose a general framework for quantum error mitigation that combines and generalizes two techniques: probabilistic error cancellation (PEC) and zero-noise extrapolation (ZNE).
Similarly to PEC, the proposed method represents ideal operations as linear combinations of noisy operations that are implementable on hardware. However, instead of assuming a fixed level of hardware noise, we extend the set of implementable operations by noise scaling. By construction, this method encompasses both PEC and ZNE as particular cases and allows us to investigate a larger set of hybrid  techniques. For example, gate extrapolation can be used to implement PEC without requiring knowledge of the device's noise model, e.g., avoiding gate set tomography. Alternatively, probabilistic error \emph{reduction} can be used to estimate expectation values at intermediate \emph{virtual} noise strengths (below the hardware level), obtaining partially mitigated results at a lower sampling cost. Moreover, multiple results obtained with different noise reduction factors can be further post-processed with ZNE to better approximate the zero-noise limit.

\end{abstract}

\maketitle

\tableofcontents

\section{Introduction}

The design of efficient techniques for reducing errors in quantum processors is an important and pressing research problem for near-term quantum computing. Standard quantum error correction codes \cite{shor1995scheme, steane1996error, calderbank1996good, aharonov2008fault, kitaev1997quantum} are theoretically known, but can require significant resources (e.g. qubits, gates) that are unavailable on near-term quantum devices \cite{preskill2018quantum}.
On the one hand, the most direct way to reduce physical errors is to improve the existing hardware, e.g., by realizing more stable qubits and less noisy operations. On the other hand, given the existing noisy hardware, large improvements can be achieved at the ``software level" by using several techniques  which have been recently called error mitigation methods \cite{temme2017error, endo2018practical, endo2021hybrid} and which are the focus of this work.

\begin{figure*}[t]
\includegraphics[width= 2 \columnwidth]{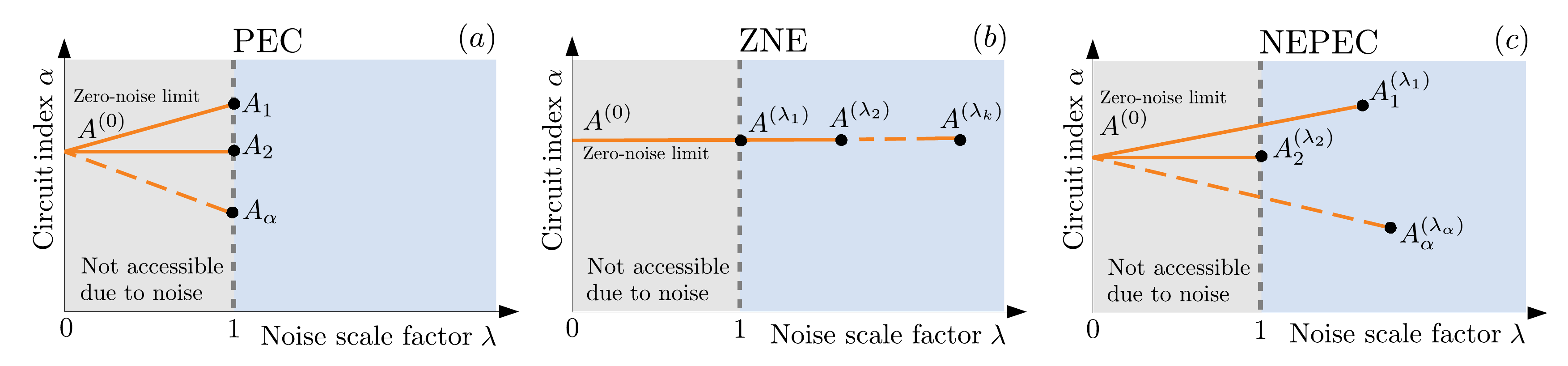} 
\caption{Pictorial representation of different error mitigation techniques. (a) In probabilistic error cancellation (PEC), an ideal expectation value is estimated from different circuits executed at the same hardware noise $\lambda=1$. (b) In zero-noise extrapolation (ZNE), an ideal expectation value is estimated from the same circuit executed at different noise levels.
(c) In NEPEC (noise-extended probabilistic error cancellation)---the error mitigation framework introduced in this work---an ideal expectation value is estimated from a general combination of different circuits evaluated
at different noise levels. In fact, the two-dimensional diagram used in this figure is a simplified representation of the more general linear combination involved in the NEPEC method. Indeed, according to  Eq.~\eqref{eq:ideal_from_nepec}, the two axes of the diagram can be multi-index arrays ($\vec{\lambda}$ and $\vec{\alpha}$), since the noise of each gate of a circuit could be scaled differently.}\label{fig:pec-zne-nepec}
\end{figure*} 

One of such methods is zero-noise extrapolation (ZNE) \cite{li2017efficient, temme2017error, kandala2019nature}, where a quantum observable is measured at different noise levels (by artificially increasing the hardware noise) and extrapolated
to the zero-noise limit. Another promising method is probabilistic error cancellation (PEC) \cite{temme2017error, endo2018practical,zhang2020error}, where ideal circuits are approximated with a Monte Carlo average over different noisy circuits.
Even if apparently different, both the ZNE and PEC methods involve the estimation of an ideal expectation value of interest from a suitable combination of noisy expectation values which are experimentally measurable. As schematically shown in Fig.~\ref{fig:pec-zne-nepec}, the main difference lies in the choice of the measured quantities: in ZNE they correspond to a fixed circuit evaluated
at different noise levels while in PEC they correspond to a set of different circuits evaluated at the same level of noise (that of the hardware). 

The aim of this work is to study a more general framework in which ideal expectation values are estimated by post-processing the results of {\it different} circuits evaluated at {\it different} noise levels. We refer to this method as NEPEC (noise-extended probabilistic error cancellation).
By construction, NEPEC includes both PEC and ZNE as limit cases. As we will show, intermediate techniques in which both degrees of freedom---the gates of the circuit and the noise level---are simultaneously exploited can be advantageous with respect to different figures of merit such as the simplicity of experimental implementation, the robustness to noise characterization errors, and the sampling cost.

Our work is complementary to the recent results of Refs.~\cite{lowe2020unified, bultrini2021unifying}, where it is shown how noise-scaling can enhance the performances of the Clifford data regression (CDR) technique introduced in Ref.~\cite{czarnik2020error} and of the virtual distillation method introduced in Refs.~\cite{koczor2020exponential, huggins2020virtual}. 
Similarly to NEPEC,  Refs.~\cite{lowe2020unified, bultrini2021unifying} combine different circuits and different noise levels. However, Refs.~\cite{lowe2020unified, bultrini2021unifying} use a  machine learning approach in which an inference model is first {\it trained} with a dataset of classically-simulable circuits and, as a second step, is applied to the circuit of interest. NEPEC instead requires the construction of (noise-aware or noise-agnostic) quasi-probability representations of ideal gates but, once such representations are determined, it can be directly applied to any circuit of interest without training.
In Ref.~\cite{piveteau2101quasiprobability}, 
a modified version of the PEC technique is proposed in which the statistical variance of the estimator can be reduced at the cost of introducing a bias error, similarly to what we study in Sec.~\ref{sec:per}. In fact, the aim of both techniques is the reduction of the negative volume of quasi-probability representations, imposing lower levels of approximation.  The main difference is that in Ref.~\cite{piveteau2101quasiprobability}, for each approximation level, a new representation must be numerically determined. In Sec.~\ref{sec:per} instead, different levels of noise reductions are analytically obtained from a suitable re-scaling of a single quasi-probability representation, exploiting the notion of canonical noise scaling introduced in Sec.~\ref{sec:canonical}. While the analytical noise scaling approach is simpler and intrinsically robust to numerical instabilities, the brute-force numerical approach of Ref.~\cite{piveteau2101quasiprobability} could achieve a lower sampling variance.
Finally, it is important to remark that the idea of combining probabilistic error reduction with ZNE was already proposed also in Ref.~\cite{cai2021multi}. In Ref.~\cite{cai2021multi}, the derivation is focused on Pauli channels and up to first order in the error probabilities. In Sec.~\ref{sec:per} we generalize this approach beyond the weak-noise limit and for arbitrary noise models.
Differences and similarities between the NEPEC framework introduced in this work and other existing techniques are summarized in Tables \ref{table} and \ref{table-literature}.

Before introducing NEPEC, we present an overview of PEC and ZNE in the next two sections, respectively. This sets the notation out and clarifies the relationship with the existing literature. The reader who is already familiar with PEC and ZNE, could directly jump to Sec.~\ref{sec:nepec} where NEPEC is defined.
In Sec.~\ref{sec:local-zne} the concept of noise-agnostic PEC is introduced, while in Sec.~\ref{sec:per} we study two techniques that involve virtual noise scaling: probabilistic error reduction (PER) and virtual ZNE. In Sec.~\ref{sec:gamma} we consider the minimal sampling cost of exact NEPEC representations, finding a no-go test for a subset of cases, and providing an example in which noise scaling is beneficial. A summary of these techniques and their key features can be found in Table \ref{table}. Finally, in Sec.~\ref{sec:conclusions} we provide concluding remarks.   

\begin{table*}
\begin{tabular}{p{5.5cm}|p{2cm}p{2.5cm}p{2.4cm}p{2.5cm}p{2.7cm}}
\bf{NEPEC technique}& \bf{Bias} & \bf{Statistical \newline variance}& \bf{Noise scaling \newline method}& 
\bf Gate  \newline tomography & \bf{Reference}\\
\hline
&&&&&\\
Probabilistic error cancellation (PEC)&
\cellcolor{g} Zero & 
\cellcolor{r} Typically large & 
\cellcolor{g} Not required & 
\cellcolor{r} Required &
Sec.~\ref{sec:pec} \\ &&&&&  Temme {\it et al.}~\cite{temme2017error} \newline Endo {\it et al.}~\cite{endo2018practical}\newline Zhang {\it et al.}~\cite{zhang2020error}\vspace{0.1 cm}\\
Zero-noise extrapolation (ZNE)&
\cellcolor{r} Potentially large & 
\cellcolor{g} Typically low & 
\cellcolor{r} Real & 
\cellcolor{g} Not required &
Sec.~\ref{sec:zne} \\ &&&&&Li {\it et al.}~\cite{li2017efficient} \newline Temme {\it et al.}~\cite{temme2017error} \newline Kandala {\it et al.}~\cite{kandala2019nature} \vspace{0.1 cm}\\
Noise-agnostic PEC \vspace{0.1 cm} &
\cellcolor{y} Intermediate & 
\cellcolor{r} Typically large & 
\cellcolor{r} Real & 
\cellcolor{g} Not required &
Sec.~\ref{sec:local-zne}\\
Probabilistic error reduction (PER) \vspace{0.1 cm}&
\cellcolor{y} Intermediate & 
\cellcolor{g} $<$ PEC variance & 
\cellcolor{g} Virtual & 
\cellcolor{r} Required &
Sec.~\ref{sec:per-a} \\ &&&&& (see also Cai~\cite{cai2021multi}) \vspace{0.1 cm}\\
Virtual ZNE \vspace{0.1 cm}&
\cellcolor{y} Intermediate & 
\cellcolor{y} Intermediate & 
\cellcolor{g} Virtual & 
\cellcolor{r} Required &
Sec.~\ref{sec:per-b}\\ &&&&& (see also Cai~\cite{cai2021multi}) \vspace{0.1 cm}\\
Exact NEPEC representations &
\cellcolor{g} Zero & 
\cellcolor{y} $\leq$ PEC variance & 
\cellcolor{r} Real & 
\cellcolor{r} Required &
Sec.~\ref{sec:gamma}\\

\end{tabular}
\caption{The table provides a qualitative summary with the key features of the different quantum error mitigation techniques reviewed (ZNE and PEC) and introduced in this work. All techniques can be seen as particular cases of the general error mitigation framework introduced in Sec.~\ref{sec:nepec}, {\it i.e.}, noise-extended probabilistic error cancellation (NEPEC). The entries of the table aim to guide the reader about typical aspects of each technique but are intentionally not rigorous and not quantitative. For example, high-order ZNE can have a large variance, while PEC applied to a short-depth circuit can have a small variance. }
\label{table}
\end{table*}

\begin{table*}
\begin{tabular}{p{1.9cm}|p{2.3cm}p{2.3cm}p{2.5cm}p{2.3cm}p{2.5cm}p{2.8cm}}
\bf Technique~~~~ & 
\bf Training & 
\bf Auxiliary \newline qubits & 
\bf Quasi-prob. \newline representations & 
\bf Noise scaling &
\bf Gate \newline tomography  & 
\bf Reference \\
\hline
&&&&&&\\
NEPEC \vspace{0.1 cm}      & \cellcolor{g} Not required     & \cellcolor{g} Not required    &  \cellcolor{r} Required                      
            & \cellcolor{y} Real or virtual  & \cellcolor{y} Optional        & This work\\
vnCDR  \vspace{0.1 cm}      & \cellcolor{r} Required         & \cellcolor{g} Not required    & \cellcolor{g} Not required
            & \cellcolor{r} Real             & \cellcolor{g} Not required    &  Lowe {\it et al.}~\cite{lowe2020unified} \\
UNITED  \vspace{0.1 cm}     & \cellcolor{r} Required         & \cellcolor{r} Required        & \cellcolor{g} Not required
            & \cellcolor{r} Real             & \cellcolor{g} Not required    & Bultrini {\it et al.}~\cite{bultrini2021unifying} \\
Bias-variance 
            & \cellcolor{g} Not required     & \cellcolor{g} Not required    & \cellcolor{r} Required
            & \cellcolor{g} Not required     & \cellcolor{r} Required        & Piveteau {\it et al.}~\cite{piveteau2101quasiprobability}\\
tradeoff    &&&&&&  \\
\end{tabular}
\caption{Qualitative table comparing the main requirements of the NEPEC framework proposed in this work to the requirements of other similar techniques existing in the literature (excluding PEC and ZNE). vnCDR stands for variable-noise Clifford data regression~\cite{lowe2020unified}. UNITED stands for UNIfied Technique for Error mitigation with Data~\cite{bultrini2021unifying}. A more detailed comparison between the different implementations of NEPEC, including the particular limit cases of PEC and ZNE, is given in Table \ref{table}.}
\label{table-literature}
\end{table*}

\section{Overview of probabilistic error cancellation}
\label{sec:pec}

We first review PEC  \cite{temme2017error, endo2018practical, takagi2020optimal}.
We consider a typical gate-based quantum computing paradigm in which the expectation value of an observable $A=A^\dag$ is evaluated after applying a unitary circuit $\mathcal U$ to $n$ qubits initially prepared in the product state $\rho_0=|0\rangle \langle 0 |^{\otimes n}$:
\begin{equation}\label{eq:a-expval}
\langle A \rangle_{\rm ideal} = {\rm tr}[A \mathcal U (\rho_0)],
\end{equation}
where the calligraphic symbol $\mathcal U$ stands for the super-operator which acts on a density matrix as $\mathcal U(\rho)=U \rho U^\dag$, where $U$ is a unitary matrix.

The circuit is assumed to be composed of a polynomial number $t$ of local unitary gates:
\begin{equation}\label{eq:u-circuit}
\mathcal U = \mathcal G_t \circ \dots \mathcal G_2 \circ \mathcal G_1,
\end{equation}
where each gate $\mathcal G_i$ typically acts non-trivially only on 1 or 2 qubits.
Now assume that, because of the hardware noise, we are actually able to apply only a set $\mathcal I$ of noisy implementable operations. An implementable operation $\mathcal O \in \mathcal I$ corresponds to a local quantum channel which is usually non-unitary and so the ideal gates of the circuit cannot be exactly implemented in a direct way. For simplicity, in this work we ignore state-preparation and measurement (SPAM) errors but, in principle, they can be taken into account in a similar fashion \cite{temme2017error, endo2018practical}.

The main idea of PEC is to represent each ideal gate $\mathcal G_i$ of the circuit as a linear combination of noisy implementable operations $\{\mathcal O_{\alpha} \}$:
\begin{equation}\label{eq:qpr}
\mathcal G_i = \sum_\alpha \eta_{i, \alpha} \mathcal O_{i, \alpha}, \quad \mathcal O_{i, \alpha} \in \mathcal I, \quad  \eta_{i, \alpha} \in \mathbb R.
\end{equation}
In principle, the index $i$ of $\mathcal O_{i, \alpha}$ in Eq.~\eqref{eq:qpr} could be dropped. However, in this work, we keep it to explicitly define a gate-dependent basis of implementable operations.

Note that this representation exists provided that: (i) the set of implementable operations is large enough to approximately or exactly represent $\mathcal G_i$ (a sufficient condition is that $\mathcal I$ forms a basis for the full space of quantum channels) and (ii) the coefficients $\eta_{i, \alpha}$ are allowed to take negative values.
Since each operation is trace-preserving we always have:
\begin{equation}\label{eq:qpr-norm}
\sum_\alpha \eta_{i,\alpha}=1,  \qquad  \gamma_i = \sum_\alpha |\eta_{i, \alpha}| \ge 1.
\end{equation}
The normalization condition implies that one can consider $\{\eta_{i,\alpha} \}$ as a {\it quasi-probability} distribution \cite{pashayan2015estimating, temme2017error} with respect to the index $\alpha$, while its one-norm $\gamma_i$ is related to the amount of {\it negativity}, i.e., the total volume of the negative coefficients. The minimum value of $\gamma_i$ is $1$ and is achieved only if all the coefficients are positive, corresponding to the special case in which $\{\eta_{i,\alpha} \}$ is a valid probability distribution. 

As we are going to see in the next subsections, the parameter $\gamma_i$ is also related to the PEC sampling cost associated to the gate $\mathcal G_i$ \cite{pashayan2015estimating, temme2017error, takagi2020optimal}. Therefore, among all the possible representations given in Eq.~\eqref{eq:qpr}, one is typically  interested in those minimizing $\gamma_i$:
\begin{align}\label{eq:gamma_opt}
\gamma_{i}^{\rm opt} = \min_{\substack{ \{ \eta_{i,\alpha} \}  \\ \{ \mathcal O_{i, \alpha} \}}}
\left[ \sum_\alpha |\eta_{i, \alpha}| \right]  \text{ s.t. Eq. \eqref{eq:qpr} holds}.
\end{align}

A detailed analysis of the optimal PEC sampling cost defined in Eq.~\eqref{eq:gamma_opt} can be found in Ref.~\cite{takagi2020optimal}.

\subsection{Error cancellation}

Now, if we first replace Eq.~\eqref{eq:qpr} into Eq.~\eqref{eq:u-circuit} and then substitute  the result into Eq.~\eqref{eq:a-expval}, we obtain an expression for the desired ideal expectation value as a linear combination of many noisy expectation values:
\begin{equation}
\langle A \rangle_{\rm ideal}= \sum_{\vec{\alpha}} \eta_{\vec{\alpha}} \langle A_{\vec{\alpha}}\rangle_{\rm noisy},  \label{eq:a-ideal-from-noisy}
\end{equation}
where:
\begin{align}
\eta_{\vec{\alpha}} &:= \eta_{t, \alpha_t} \dots \eta_{2, \alpha_2} \eta_{1,\alpha_1},\\
\langle A_{\vec{\alpha}}\rangle_{\rm noisy} &:=  {\rm tr}[A \Phi_{\vec{\alpha}}(\rho_0)], \\
\Phi_{\vec{\alpha}} &:= \mathcal O_{t, \alpha_t} \circ \dots \circ \mathcal O_{2, \alpha_2} \circ \mathcal O_{1, \alpha_1}. \label{eq:noisy-circuit}
\end{align}
By linearity of the sum, the coefficients $\eta_{\vec{\alpha}}$ form again a quasi-probability distribution
for the global circuit whose one-norm $\gamma$ is the product of those of the gates:
\begin{equation}\label{eq:full-qpr-norm}
\sum_{\vec \alpha} \eta_{\vec{\alpha}}=1,  \qquad  \gamma = \sum_{\vec{\alpha}} |\eta_{\vec \alpha}| = \Pi_i \gamma_i.
\end{equation}

In principle, by running all the noisy circuits $\Phi_{\vec{\alpha}}$ and evaluating all the corresponding expectation values $\langle A_{\vec{\alpha}} \rangle$, one can exactly compute the ideal result $\langle A \rangle_{\rm ideal}$. However, the number of terms in Eq.~\eqref{eq:a-ideal-from-noisy} grows exponentially with respect to the number of gates $t$ and, in most practical situations, this approach is unfeasible.  
A possible solution to avoid this issue is to replace the sum in Eq.~\eqref{eq:a-ideal-from-noisy} with a Monte Carlo approximation \cite{pashayan2015estimating, temme2017error, endo2018practical, takagi2020optimal}.  

\subsection{Monte Carlo estimation}

Let us define the probability distribution $p_{i}(\alpha)=|\eta_{i, \alpha}|/\gamma_i$ associated to the quasi-probability representation of the gate $\mathcal G_i$. It is easy to check that we can always re-write Eq.~\eqref{eq:qpr} as:
\begin{equation}\label{eq:qpr-p}
\mathcal G_i =  \sum_\alpha p_{i}(\alpha) \gamma_i {\rm sign}(\eta_{i, \alpha}) \mathcal O_{i, \alpha}.
\end{equation}
If we sample an index $\alpha$ at random from the distribution $p_{i}(\alpha)$, we obtain a random variable $\hat \alpha$ such that
\begin{equation}\label{eq:G-estimator}
\hat{\mathcal G}_{i} = \gamma_i {\rm sign}(\eta_{i, \hat{\alpha}}) \mathcal O_{i, \hat \alpha},
\end{equation}
is an unbiased estimator for the ideal gate $\mathcal G_i$, i.e., $\mathcal G_i= \mathbb E \{ \hat{\mathcal G_i}\}$ where
the average $\mathbb E(\cdot)$ is with respect to the sampling distribution $p_{i}(\alpha)$.

If, for each gate $\mathcal G_i$ of the circuit described in Eq.~\eqref{eq:u-circuit}, we independently sample an index $\hat \alpha$ corresponding to the noisy operation $\mathcal O_{i, \hat \alpha}$, we can define an unbiased estimator for the full circuit as:

\begin{align}\label{eq:u-estimator}
\hat{ \mathcal U} 
&=   \hat{\mathcal G}_t \circ \dots \hat{\mathcal G}_2 \circ \hat{\mathcal G}_1= \gamma \sigma_{\vec{\hat{\alpha}}} \Phi_{\vec{\hat{\alpha}}},
\end{align}
where $\Phi_{\vec{\hat{\alpha}}}$ is the sampled noisy circuit, $\gamma$ is the constant introduced in Eq.~\eqref{eq:full-qpr-norm} and $\sigma_{\vec{\hat{\alpha}}}= \Pi_i {\rm sign}(\eta_{i, \hat \alpha})={\rm sign}(\eta_{\vec{\hat \alpha}})$.
In other words, by sampling individual gates one-by-one, we are effectively sampling from the global quasi-probability representation 
of the full circuit such that $\mathcal U=\mathbb E \{\hat{\mathcal U} \} $. 
This directly implies that by measuring the observable $A$ on the sampled circuits, one can estimate the ideal expectation value with a Monte Carlo sampling average over noisy expectation values:

\begin{align}\label{eq:a-ideal-from-samples}
\langle A \rangle_{\rm ideal}= \mathbb E \left\{ {\rm tr}[A \hat{\mathcal{U}}(\rho_0)] \right\} 
=   \mathbb E \left\{\gamma \sigma_{\vec{\hat \alpha}} \langle A_{\vec{\hat{\alpha}}} \rangle_{\rm noisy} \right\}.
\end{align}

Differently from Eq.~\eqref{eq:a-ideal-from-noisy} which is exact but contains an exponential number of terms, the right-hand-side of Eq.~\eqref{eq:a-ideal-from-samples} can be approximated by averaging over a finite number of samples. How many samples are necessary to approximate $\langle A \rangle_{\rm ideal}$ up to a precision $\delta$?
From the standard (classical) theory of Monte Carlo sampling it can be shown that \cite{pashayan2015estimating, takagi2020optimal}:  
\begin{equation}
\text{\#  of samples}  \propto \frac{\gamma^2}{\delta^2}.
\end{equation}
This gives a clear operational meaning to the parameter $\gamma$: the larger the negativity of the quasi-distribution the higher the PEC sampling cost.
Since $\gamma=\Pi_i \gamma_i$, its value scales exponentially with respect to the number of gates. Nonetheless, PEC can still be very advantageous with the medium-size circuits which can run on near-term quantum computers \cite{zhang2020error}.

\section{Overview of zero-noise extrapolation}
\label{sec:zne}

Zero-noise extrapolation \cite{li2017efficient, temme2017error, kandala2019nature} is an error mitigation method that relies on the
assumption that the noise of the physical hardware can be artificially increased beyond the base
level such that, after measuring an expectation value for different strengths of the noise, it is possible to extrapolate the zero-noise limit.

More precisely, it is assumed that the strength of the noise can be quantified by some parameter $\Gamma$ (e.g. the decoherence rate of some noise channel) and that it is possible to scale the noise by a dimensionless parameter $\lambda \ge 1$ such that
the effective noise channel acting on the system has a larger strength $\Gamma' = \lambda \Gamma$. Importantly, differently from PEC, in this case it is not necessary to know the details of the noise model. In fact, there are many situations in which the noise level $\Gamma$ is unknown and/or the associated noise channel is hard to characterize, nonetheless, it is still possible to correctly scale $\Gamma$ by some given factor $\lambda$. For example, the pulse-stretching technique introduced in \cite{temme2017error} and several gate-level methods proposed in \cite{giurgica2020digital, dumitrescu2018cloud, he2020resource} allow to effectively  scale the hardware noise, without actually ``knowing" it (within the limits and assumptions of these methods).

We now try to express ZNE with a formalism that is as similar as possible to the one previously used for describing PEC. We consider an ideal expectation value $\langle A \rangle_{\rm ideal}$ that we would like to infer from a set of noisy expectation values $\langle A \rangle_{\rm noisy}^{(\lambda)}$ evaluated at different values of the noise scale factor  $\lambda \ge 1$:
\begin{align}
\langle A \rangle_{\rm noisy}^{(\lambda)} &:=  {\rm tr}[A \mathcal U^{(\lambda)}(\rho_0)], \\
\mathcal U^{(\lambda)} &:= 
\mathcal G_t^{(\lambda)} \circ \dots  \circ \mathcal G_2^{(\lambda)} \circ \mathcal G_1^{(\lambda)},
\end{align}
where $\mathcal G_i^{(\lambda)}$ represents the noise-scaled implementation of the ideal unitary gate $\mathcal G_i =\mathcal G_i^{(\lambda=0)} $.
If a polynomial fit \cite{giurgica2020digital} is used to evaluate the zero-noise limit  (Richardson extrapolation \cite{temme2017error} is a special case), the ideal result can always be expressed as linear combination of the measured noisy expectation values:
\begin{align}\label{eq:a-ideal-from-noisy-ZNE}
\langle A \rangle_{\rm ideal}  = \sum_{\lambda \in S} \eta_\lambda \langle A \rangle_{\rm noisy}^{(\lambda)} + \Delta \, ,
\end{align}
where $S=\{\lambda_1, \lambda_2, \dots, \lambda_m \}$ is the chosen set of $m$ noise scale factors and where $\eta_\lambda$ are real coefficients which are completely determined by $S$ and by the degree of the polynomial but are independent of the measured
results (because a polynomial fit is a linear regression problem)
\footnote{
This is valid for a basic polynomial regression. For more advanced inference algorithms, e.g.~those taking into account different uncertainties for each data point, the extrapolation coefficients can be data-dependent.
}
. Differently from the exact linear combination introduced in Eq.~\eqref{eq:a-ideal-from-noisy}, in ZNE the result is subject to a some bias error $\Delta$, which depends on the amount of noise $\Gamma$ and on the quality of the extrapolation model.

It is easy to check that, also in this case,
$\{ \eta_\lambda \, |\, \lambda \in S\}$ can be interpreted as a
quasi-probability distribution:
\begin{equation}\label{eq:qpr-norm-zne}
\sum_{\lambda \in S} \eta_{\lambda}=1,  \qquad  \gamma = \sum_{\lambda \in S} |\eta_{\lambda}| \ge 1.
\end{equation}
Similarly to PEC, the parameter $\gamma$ quantifies how much the statistical uncertainty on the expectation values $ \langle A \rangle_{\rm noisy}^{(\lambda)}$ gets amplified by the linear combination in Eq.~\eqref{eq:a-ideal-from-noisy-ZNE} and so how large is the mitigation overhead. For example, in the case of Richardson extrapolation, $\gamma$ scales exponentially with respect to the size of $S$ \cite{giurgica2020digital, he2020resource} which in practice implies that only a few noise scale factors must be used to avoid numerical instabilities. In general, the number of terms in the linear combination is small and one can directly measure all the noisy expectation values. This means that, in the case of ZNE, it is not necessary to use any probabilistic Monte Carlo sampling.

To summarize, in both the PEC and ZNE methods, an ideal expectation value is represented as an average over noisy expectation values with respect to a quasi probability distribution. The only difference lies in the domain of the  distribution as shown in Fig.\ref{fig:pec-zne-nepec}. Can we consider the two domains of PEC and ZNE as particular cases of a more general configuration space?
This is the main idea proposed in this work and we formalize it in the next section.

\section{NEPEC: Noise-extended probabilistic error cancellation}
\label{sec:nepec}

In this section we introduce NEPEC (noise-extended probabilistic error cancellation), a generalization of PEC in which the set of implementable operations is extended by noise scaling (see Fig.~\ref{fig:pec-zne-nepec}).

We begin with the set of implementable operations $\mathcal I$ which appears in Eq.~\eqref{eq:qpr} and which is at the basis of PEC.
Let us also assume that for each  operation $\mathcal O \in  \mathcal I$,  one can scale the noise by a factor  $\lambda \ge 1$, obtaining a noise-scaled operation $\mathcal O^{(\lambda)}$. In practice this implies that, thanks to noise scaling, we have at our disposal an extended set of implementable operations:
\begin{equation}\label{eq:extended-implementable-ops}
\tilde{\mathcal I} = \{\mathcal O^{(\lambda)}\, | \,  \mathcal O \in \mathcal I,\; \lambda \ge 1 \}.
\end{equation}
By construction, $\tilde{\mathcal I}$ includes $\mathcal I$ as a subset.
A noise-scaled implementable operation $\mathcal O^{(\lambda)}$ can be expressed as:
\begin{equation}\label{eq:o-lambda}
\mathcal O^{(\lambda)}:= \mathcal G^{(\lambda)} = \mathcal E^{(\lambda)} \circ \mathcal G,
\end{equation}
where $\mathcal G$ is some ideal unitary operation and $\mathcal E^{(\lambda)}$ is some noise channel such that  $\mathcal E^{(1)}$ corresponds to the hardware base noise and $\mathcal E^{(0)} = {\rm Id}$. For example, if a gate $\mathcal G$ acting on $k$ qubits is affected by a depolarizing channel
$\mathcal D_p$ with error probability $p$ \cite{takagi2020optimal}, the associated noise-scaled operation can be defined as in Eq.~\eqref{eq:o-lambda} with:
\begin{equation}\label{eq:depo-scaled}
\mathcal E^{(\lambda)}=\mathcal D_{\lambda p} = (1 -  \lambda p) {\rm Id} + \lambda p \sum_{\mathcal P \in P_k} \frac{\mathcal P}{4^{k} - 1}, \quad \lambda \in [1, \lambda_{\rm max}],
\end{equation}
where $P_k$ is the set of all Pauli strings of length $k$ with the exclusion of the identity string ${\rm Id}_1 \otimes {\rm Id}_2 \otimes .... \otimes {\rm Id}_k$ and $\lambda_{\rm max}$ is the maximum
scale factor. If we impose $\lambda p$ to be a valid probability we obtain $\lambda_{\rm max}=p^{-1}$, while
if we more realistically require that the maximum scaling is achieved when the output state is completely
mixed we get $\lambda_{\rm max}= (1 -4^{-k})p^{-1}<p^{-1}$.\\

In practice, noise scaling can be experimentally achieved by acting on the physical control pulses \cite{temme2017error} or by acting on the circuit at a gate-level  \cite{dumitrescu2018cloud, he2020resource, giurgica2020digital}. For example, we anticipate that in the simulation shown in Fig.~\ref{fig:noise-agnostic-pec} (presented in the next section), we will apply {\it unitary folding}   \cite{giurgica2020digital} to approximately scale the noise with odd integers values of $\lambda$. According to this method, instead of using Eq.~\eqref{eq:o-lambda}, a noise-scaled operation associated to a gate $\mathcal G$ is defined as:
\begin{align}\label{eq:unitary-folding}
&\mathcal O^{(\lambda)} := \mathcal G^{(\lambda)}= \mathcal G^{(1)}\circ 
\left [\mathcal G^{\dag (1)} \circ \mathcal G^{(1)}\right]^{(\lambda - 1)/2},  \nonumber \\ 
&\lambda=1, 3, \dots ,
\end{align}
where $\mathcal G^{(1)}$, corresponds to the gate $\mathcal G$ executed at the hardware base noise (i.e., at $\lambda=1$). More details on the practical advantages and limitations of unitary folding can be found in Refs. \cite{giurgica2020digital, larose2020mitiq}. Even if the final results may depend on the choice of the noise scaling method, the abstract formulation of the NEPEC technique is actually independent from this choice.

If we expand an ideal gate $\mathcal G_i$ of a given circuit in the new basis $\tilde {\mathcal I}$, we obtain:

\begin{equation}\label{eq:extended-qpr}
\mathcal G_i = \sum_{\alpha, \lambda} \eta_{i, \alpha, \lambda} \mathcal O_{i, \alpha}^{(\lambda)}, \quad \mathcal O_{i, \alpha}^{\lambda} \in \tilde{\mathcal I}, \quad  \eta_{i, \alpha, \lambda} \in \mathbb R,
\end{equation}
which is the natural generalization of Eq.~\eqref{eq:qpr} to a noise-extended basis. The associated mitigation cost is given by the one-norm of the quasi-probability distribution $\{\eta_{i,\alpha,\lambda}\}$, i.e., $\gamma_i = \sum_{\alpha, \lambda}|\eta_{i, \alpha, \lambda}| \ge 1$.

Similarly to PEC and ZNE, our goal is to approximate an ideal expectation value as a linear combination 
of noisy expectation values associated to implementable circuits:
\begin{align}
\langle A \rangle_{\rm ideal} &= \sum_{\vec{\alpha},  \vec \lambda} \eta_{\vec{\alpha},  \vec \lambda}  
\langle A_{\vec \alpha} \rangle_{\rm noisy}^{(\vec \lambda)}  \label{eq:ideal_from_nepec}\\
\langle A_{\vec \alpha} \rangle_{\rm noisy}^{(\vec \lambda)} &:=  {\rm tr}[A  \Phi_{\vec \alpha}^{(\vec \lambda)}(\rho_0)], \\
\Phi^{(\vec \lambda)}_{\vec{\alpha}} &:= 
\mathcal O_{t, \alpha_t}^{(\lambda_t)} \circ \dots  \circ  \mathcal O_{2, \alpha_2}^{(\lambda_2)} \circ \mathcal O_{1, \alpha_1}^{(\lambda_1)},
\end{align}
where $\mathcal O^{(\lambda)}_{i, \alpha} \in \tilde{\mathcal I}$ is $i$-th noisy gate of the circuit, $\alpha$ enumerates different implementable operations and $\lambda$ is the noise scale factor.

Since in NEPEC we have two degrees of freedom---the choice of gates and their noise scale factors---the set of coefficients $\{ \eta_{\vec \alpha, \vec \lambda} \}$ should be considered as a quasi-distribution with respect to the pair of indices  $(\vec \alpha, \vec{\lambda})$:

\begin{equation}\label{eq:full-qpr-norm-nepec}
\sum_{\vec \alpha, \vec{\lambda}} \eta_{\vec{\alpha}, \vec \lambda}=1,  \qquad  \gamma = \sum_{\vec{\alpha}, \vec{\lambda}} |\eta_{\vec \alpha, \vec \lambda}| .
\end{equation}

In most practical situations, it is not possible to measure all the terms in Eq.~\eqref{eq:ideal_from_nepec}  because their number is too large. In these cases, one can sample a random pair of indices $(\vec {\hat \alpha}, \vec{{\hat \lambda}})$  from the probability distribution $p_{\vec \alpha, \vec \lambda} = |\eta_{\vec \alpha, \vec \lambda}| /\gamma$, such that the ideal expectation value can be estimated probabilistically as an average over many samples:

\begin{align}\label{eq:a-ideal-from-samples-nepec}
\langle A \rangle_{\rm ideal}= 
\mathbb E \left\{\gamma \sigma_{\vec{\hat \alpha}, \vec {\hat \lambda}} \langle A_{\vec{\hat{\alpha}}}^{(\vec {\hat \lambda})} \rangle_{\rm noisy} \right\},
\end{align}
where $\sigma_{\vec{\hat \alpha}, \vec {\hat \lambda}}= {\rm sign}(\eta_{\vec{\hat \alpha}, \vec {\hat \lambda}})$. This is the NEPEC version of  Eq.~\eqref{eq:a-ideal-from-samples} and, also in this case, the sampling cost scales as $\gamma^2$.

In Eq.~\eqref{eq:ideal_from_nepec}, the sum is over the two variables $\vec{\alpha}$ and $\vec \lambda$. If we fix $\lambda_j=1$, we re-obtain the PEC linear combination given in Eq.~\eqref{eq:a-ideal-from-noisy}. If we instead fix the choice of the noisy gates to match that of the ideal circuit and impose a uniform scale factor $\lambda_j=\lambda$, the sum over $\vec \alpha$ disappears and we recover the ZNE linear combination introduced in Eq.~\eqref{eq:a-ideal-from-noisy-ZNE}.
A graphical representation showing how PEC and ZNE are particular cases of NEPEC is given in Fig.~\ref{fig:pec-zne-nepec}.

Is there any advantage in using the more general NEPEC framework compared to the
particular limit cases represented by PEC and ZNE? The answer to this question depends on the considered figure of merit and so it is better to address more specific questions:
\begin{enumerate}

\item Can NEPEC be used without the full tomographic knowledge of the gates which  is instead necessary in PEC? We address this question in Sec.~\ref{sec:local-zne}.
\item Can NEPEC be used to effectively implement virtual noise scaling methods for ZNE? We address this question in Sec.~\ref{sec:per}.
\item Can NEPEC allow for a smaller sampling cost compared to PEC? We address this question in Sec.~\ref{sec:gamma}.

\end{enumerate}

In the next sections we show, with theoretical arguments and with explicit numerical examples, that all the previous questions can have a positive answer, even though this depends on the specific circuit, gate set, and device under consideration. The techniques introduced hereafter are qualitatively compared to ZNE and PEC in Table \ref{table}.

\section{Extrapolating gates for noise-agnostic PEC}
\label{sec:local-zne}

In the previous section we presented the general formulation of NEPEC, introducing the quasi-probability representations for expectation values [Eqs.~\eqref{eq:ideal_from_nepec} and \eqref{eq:a-ideal-from-samples-nepec}], and for individual gates (Eq.~\eqref{eq:extended-qpr}). 
However, among all the possible representations which could be used, what are the good ones for practical applications?
Are there specific NEPEC representations that have some good physical motivation or that are practically easier to implement?

In this section we consider a particular technique where the quasi-probability representation for individual gates is inspired by ZNE while the full circuit is sampled according to the standard PEC algorithm. The main advantage of this approach is the possibility of obtaining an approximate probabilistic error   cancellation  technique which is {\it noise-agnostic}, i.e., which does not require the full characterization of the noise model or the full tomography of the noisy gate set.

The idea is to represent each ideal gate $\mathcal G_i$ of a circuit as a linear combination of the same noisy gate executed at different noise scale factors $\lambda \ge 1$. In practice we drop the $\alpha$ index which appears in the NEPEC representation of a generic gate given in Eq.~\eqref{eq:extended-qpr} and keep only the sum over a finite set of noise scale factors $S_i=\{\lambda_1, \lambda_2, \dots \}$:

\begin{equation}\label{eq:local-zne-qpr}
\mathcal G_i = \sum_{\lambda \in S_i} \eta_{i,\lambda} \, \mathcal G_i^{(\lambda)}
+ \Delta_i,  \quad  \eta_{i, \lambda} \in \mathbb R,
\end{equation}
where we also add a small bias error---the super-operator $\Delta_i$ in \eqref{eq:local-zne-qpr}---since the restricted noisy basis may not allow for an exact representation. 

Note that Eq.~\eqref{eq:local-zne-qpr} 
could be considered as a kind of ``extrapolation" of the noisy gate $\mathcal G_i^{(\lambda)}$ to the zero-noise limit $\mathcal G_i^{(\lambda=0)}$. Differently from standard ZNE, where the extrapolation is applied to some scalar expectation value, here we are instead extrapolating the super-operator of a single gate with the aim of obtaining a good quasi-probability representation.

\begin{figure}
     \includegraphics[width=1.05 \columnwidth]{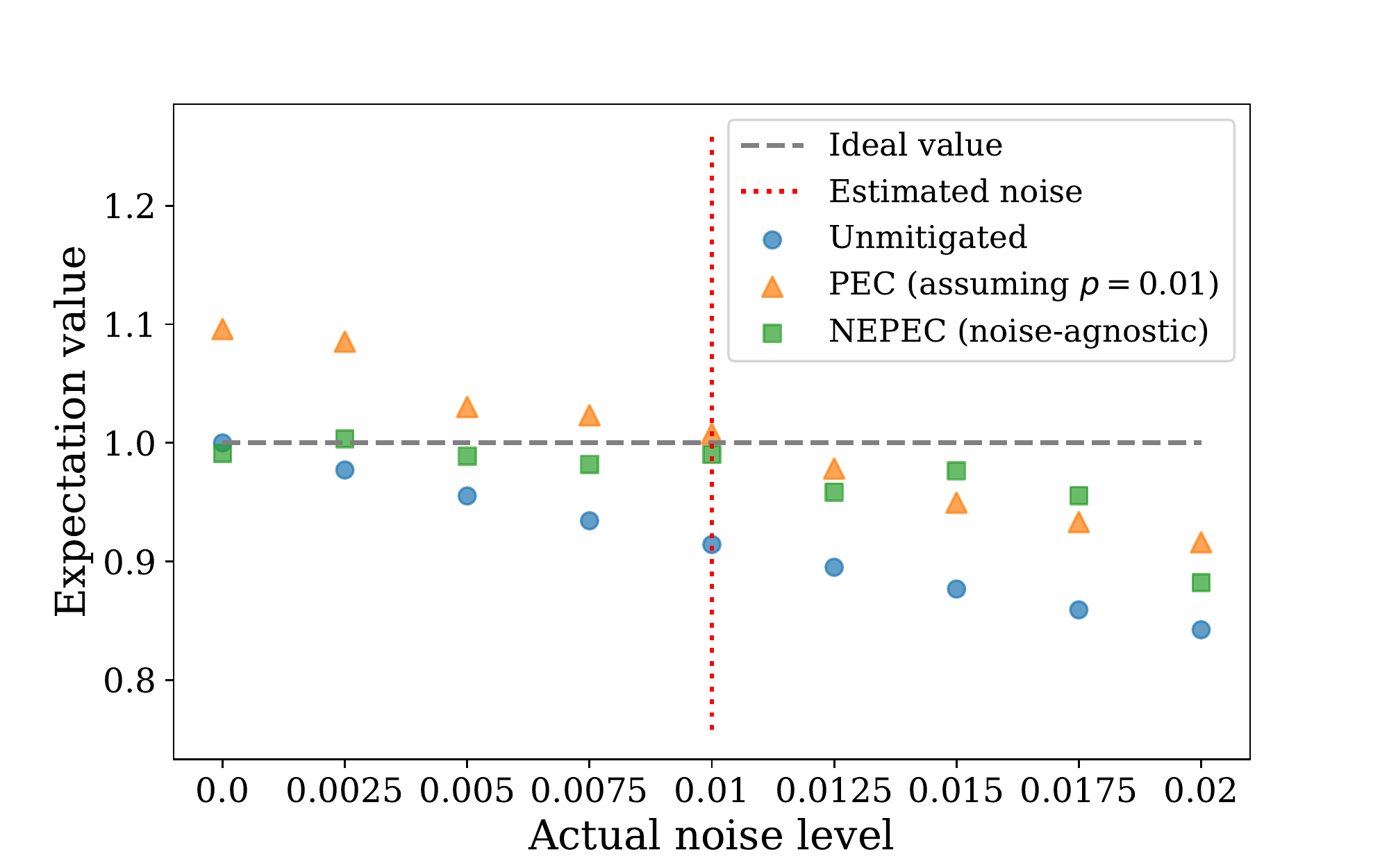}
    \caption{
    The expectation value of the observable  $A=|0\rangle\langle0|$ is estimated with different techniques  
    for a single-qubit randomized benchmarking circuit of depth $14$ such that $\langle A \rangle_{\rm ideal}=1$. The PEC results are based on quasi-probability representations built assuming a fixed estimated noise level of $p=0.01$. When this assumption matches the actual noise level of the hardware, PEC achieves an optimal cancellation of the noise. On the other hand, when the actual noise of the system is different from the assumed value (noise characterization error), PEC results are more biased than NEPEC results.
    The NEPEC results are obtained using the noise-agnostic representations defined according to  Eq.~\eqref{eq:local-zne-qpr} with scale factors $S=\{\lambda_1, \lambda_2\}=\{1, 51\}$ and coefficients given by Eq.~\eqref{eq:richardson-qpr}. Unitary folding is used for digitally scaling the noise as described in Eq.~\eqref{eq:unitary-folding} without assuming a specific noise model or noise level.  All the noisy expectation values are directly evaluated from simulated density matrices and therefore shot noise is absent in this figure (corresponding to the limit of infinite shots). The random  fluctuations of the PEC and NEPEC points are due to the finite number (5000) of Monte Carlo samples. For the numerical simulation of this example, we used the error mitigation software package Mitiq \cite{larose2020mitiq}. }
    \label{fig:noise-agnostic-pec}
\end{figure}

Crucially, the scalar coefficients $\eta_{i, \lambda}$ can be determined in exactly the same way as in standard ZNE. More precisely, for any polynomial model (including Richardson extrapolation), the coefficients ${\eta_{i, \lambda}}$ depend only on the set of noise scale factors $S_i$. This means that, given the extrapolation order and the set $S_i$, we can compute the coefficients ${\eta_{i, \lambda}}$ with the same standard methods which are often used for ZNE.
For example, in the case of Richardson extrapolation, the explicit formula for the coefficients is \cite{giurgica2020digital}:

\begin{equation}\label{eq:richardson-qpr}
\eta_{i,\lambda}=\prod_{\substack{\lambda' \in S_i \\ \lambda' \neq \lambda}} \frac{\lambda'}{\lambda' - \lambda}, \quad \lambda \in S_i.
\end{equation}
Alternatively, only in the case in which one has the full  tomographic knowledge of the noise scaled gates, one could directly  optimize the coefficients ${\eta_{i, \lambda}}$ in Eq.~\eqref{eq:local-zne-qpr} to better represent the ideal gate.

Once we have the quasi-probability representation described in Eq.~\eqref{eq:local-zne-qpr} for each gate of an ideal  circuit, 
we can estimate expectation values via the usual probabilistic error cancellation approach.  
In practice one has to stochastically modulate the effective noise along the circuit by sampling a noise scale factor for each gate according to the probability distribution $p_{i}(\lambda)=|\eta_{i, \lambda}|/\gamma_i$. Eventually, to estimate the ideal result one should combine all the measured samples according to Eq.~\eqref{eq:a-ideal-from-samples-nepec}.

Compared to the standard decomposition used in PEC and introduced in Eq.~\eqref{eq:qpr}, in order to use the specific NEPEC representation given in Eq.~\eqref{eq:local-zne-qpr}, one needs some direct or indirect way of scaling the noise of individual gates. On the other hand, the main advantage of this method is that it does not require the full knowledge of the noise model which would instead be necessary to evaluate Eq.~\eqref{eq:qpr}.

A simple example demonstrating noise-agnostic error mitigation with NEPEC is shown in Fig.~\ref{fig:noise-agnostic-pec} for a single-qubit circuit subject to different levels of depolarizing noise.
In Fig.~\ref{fig:noise-agnostic-pec}, one can also observe that standard PEC (with fixed gate representations) is quite sensible to noise characterization errors ({\it i.e.} the mismatch between the assumed and the actual noise-models). On the contrary, exploiting the noise-independent representations defined in Eq.~\eqref{eq:local-zne-qpr}, NEPEC is by construction more robust with respect to the noise level and to the noise type.
\\

\noindent {\bf Note:} Instead of considering the unitary operator $\mathcal G_i$ in  Eq.~\eqref{eq:local-zne-qpr} as the $i$-th gate of the circuit, one may also associate it to the $i$-th layer or to the $i$-th sub-circuit. The proposed  procedure would be exactly the same, with the only constraint of using a uniform noise scale factor for each $i$-th part of the circuit. Depending on the noise scaling method, acting at the level of layers may be more practical with respect to scaling the noise of individual gates. For example, it is probably simpler to apply the pulse-stretching \cite{temme2017error, kandala2019nature} technique layer-wise instead of gate-wise. \\

\subsection{Reducing the sampling cost of gate extrapolation}
 
The sampling cost parameters $\gamma_i$ of the quasi-probability representation defined in Eq.~\eqref{eq:local-zne-qpr} can be quite large compared to the PEC representation of Eq.~\eqref{eq:qpr}  and this fact can be a practical limitation for large circuits. We propose two solutions to reduce the sampling cost in the case of large and highly noisy circuits:
\begin{enumerate}
\item[1] 
Considering $\mathcal G_i$ as the unitary matrix which represents a large block of the circuit instead of a single gate or of a thin layer of gates.
\item[2] Applying large noise scaling to increase the distance between the different elements of $S_i$ (the set of noise scale factors).
\end{enumerate}

An example of the first option is to divide a long circuit into a few (say 2 or 3)  parts of approximately equal size. Each part could be represented as in Eq.~\eqref{eq:local-zne-qpr}. In this setting, the full circuit is expressible as a linear combination of a limited number of terms which can be easily measured.

The second option instead is motivated by the fact that, the more the values of $\lambda$ are different, the smaller the one-norm of the extrapolation coefficients. For example, for a linear extrapolation with scale factors  $S_i=\{1, \lambda_2\}$, the two extrapolation coefficients would be $\eta_{i,1} = \lambda_2/(\lambda_2- 1)$ and $ \eta_{i,\lambda_2}=1/(1- \lambda_2)$, corresponding to a mitigation cost of $\gamma_{i}=(1 + \lambda_2)/(\lambda_2 - 1)$ which tends to one for a large $\lambda_2$. In practice, however, there is usually a trade-off for the optimal amount of noise scaling: large noise scaling values are convenient for reducing the sampling variance, but too large noise scaling may give a bad extrapolation bias (large $\Delta$ in Eq.~\eqref{eq:local-zne-qpr}). Moreover, depending on how noise scaling is defined, there can be a maximum value of lambda $\lambda$ above which it is impossible to scale the noise. For example, for the noise-scaled depolarizing channel $\mathcal D_{\lambda p}$ acting on $k$ qubits defined in Eq.~\eqref{eq:depo-scaled}, a physically motivated maximum value is $\lambda_{\rm max}=(1- 4^{-k}) p^{-1}$, where $p$ is the base error rate of the hardware. In this case, linear extrapolation of a single-qubit gate would give a mitigation cost of:
\begin{equation} \label{eq:optimal_zne_cost_nepec}
\gamma_i =\frac{1 + \lambda_{\rm max}}{\lambda_{\rm max} -1} =  \frac{1 + \epsilon}{1 - \epsilon}, \quad \epsilon = (4/3) p.
\end{equation}
This is larger than the PEC optimal cost $\gamma_{\rm opt} = (1 + \epsilon/2)(1- \epsilon)$ \cite{takagi2020optimal},
consistent with the no-go result of Sec.~\ref{sec:gamma}.
Interestingly, if we instead set $\lambda_{\rm max}= p^{-1}$, we obtain $\gamma_i = (1 + p)/(1 - p)$
which is slightly smaller than the optimal PEC cost of Ref.~\cite{takagi2020optimal}. However one should take into account that, for $\lambda  >  (1- 4^{-k}) p^{-1}$, the corresponding noise-scaled depolarizing channel may be impossible to implement on hardware even if mathematically well-defined. For this reason, Eq.~\eqref{eq:optimal_zne_cost_nepec} is a more realistic and prudent estimate of the actual error mitigation cost.

\section{Probabilistic error reduction and virtual ZNE} \label{sec:per}

In this section we show how one can use a probabilistic sampling approach for indirectly  implementing a  ``virtual" noise scaling process instead of aiming for a complete cancellation of errors. Differently from common noise scaling methods like pulse-stretching or unitary folding, virtual noise scaling can be used to effectively reduce the noise below the hardware level instead of scaling it up.
Similarly to the general  NEPEC framework discussed in Sec.~\ref{sec:nepec}, also in this section we extend  the space of implementable operations to arbitrary (virtual) noise levels and we show how this can be useful for reducing the statistical variance of the mitigated results.

We comment that a similar notion of probabilistic noise reduction was recently proposed also in Ref.~\cite{cai2021multi}.

\begin{figure}[t]
    \centering
    \includegraphics[width=1 \columnwidth]{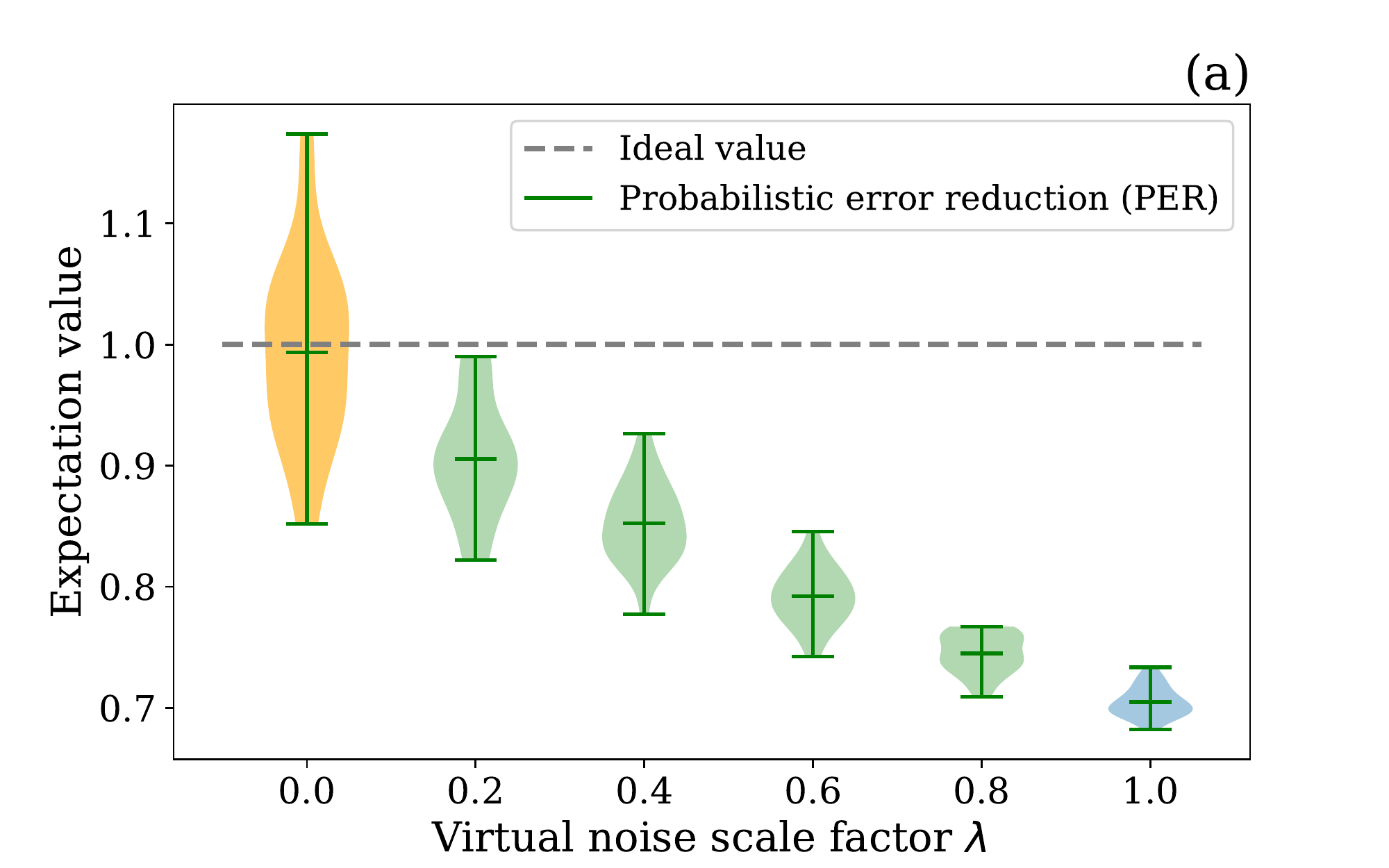}
    \vspace{-0.2 cm}
    \includegraphics[width=1 \columnwidth]{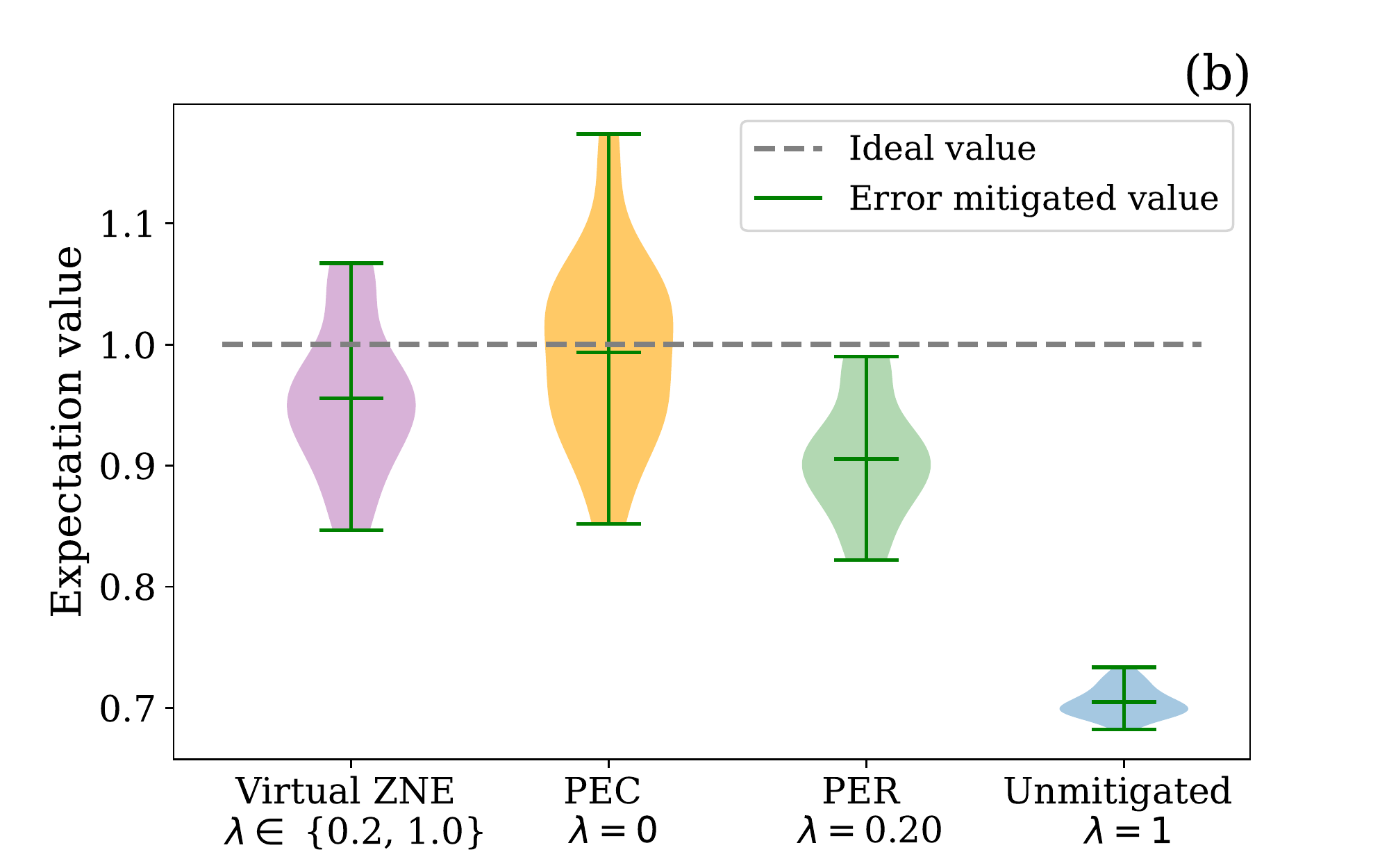}
    \caption{
    (a) Probabilistic error reduction by virtual noise scaling. A single-qubit randomized-benchmarking circuit composed of 46 gates is simulated assuming a depolarizing noise model with $p=0.015$. The expectation value of $A=|0\rangle\langle 0|$ is estimated at different values of the noise scale factor $\lambda$ via PER according to Eq.~\eqref{eq:qpr-for-per} and, more specifically, \eqref{eq:per_depolarizing}. For increasing values of $\lambda$, the  results continuously interpolate between PEC ($\lambda=0$) and no mitigation ($\lambda=1$), while the associated statistical variance gets reduced.
    (b) Error mitigation by linear zero-noise extrapolation with virtual scale factors $S=\{\lambda_1, \lambda_2\}=\{0.2, 1\}$, compared to standard PEC and to PER (with $\lambda=0.2$). We observe that all error mitigation methods improve the unmitigated result. As expected, virtual ZNE and PER are on average more biased than PEC, however,  their statistical uncertainty can be significantly smaller.
    For each noise scale factor, $5\times10^4$ measurements (shots) have been used. For each shot, a different circuit was sampled according to PER. To estimate the statistical distribution, the samples have been divided into $25$ batches of $2 \times 10^3$ elements each. The ``violin plots" show the statistical distributions with respect to  independent batches. The central horizontal segment in each ``violin" corresponds to the mean of all the samples, which is equal to the mean of the batches. For the numerical simulation of this example, we used the error mitigation software package Mitiq \cite{larose2020mitiq}.}
    \label{fig:virtual-zne}
\end{figure}

\subsection{Probabilistic error reduction}
\label{sec:per-a}

For noise scale factors $\lambda$ larger than 1 (noise amplification), probabilistic noise scaling is not new. Indeed the probabilistic application of Pauli gates for amplifying the noise, was already used in one of the first experimental applications of ZNE~\cite{li2017efficient}.
In this work instead we explore the possibility of virtually scaling down the 
noise in the interval $\lambda \in [0, 1]$ and we refer to this technique as probabilistic error reduction (PER), which can be considered both as noise scaling method and as an error mitigation method. 

Given a generic definition of noise scaling (see Eq.~\eqref{eq:o-lambda}), PER can be obtained generalizing the PEC quasi-probability representation given in Eq.~\eqref{eq:qpr}. Specifically, we replace the unitary gate on the l.h.s. of Eq.~\eqref{eq:qpr} with a non-unitary channel corresponding to the $i$-th noise-scaled gate $\mathcal G_i^{(\lambda)} $ of the circuit of interest:

\begin{align}\label{eq:qpr-for-per}
 \mathcal G_i^{(\lambda)}  &= \sum_\alpha \eta_{i, \alpha}(\lambda) \, \mathcal O_{i, \alpha},  \quad 
\mathcal O_{i, \alpha} \in \mathcal I, \\
\sum_\alpha &  \eta_{i, \alpha}(\lambda) = 1 \quad  \forall \lambda, \qquad \gamma_i(\lambda) := \sum_\alpha |\eta_{i, \alpha}(\lambda)|. \nonumber
\end{align}

Explicit quasi-probability representations for the particular case of depolarizing noise are derived in subsection \ref{subsec:depo-per} and a more general scenario is considered subsection \ref{sec:canonical}. An example of PER is shown in Fig.~\ref{fig:virtual-zne}(a), where an expectation value is estimated at different virtual noise scale factors, interpolating between the unmitigated result ($\lambda=1$) and PEC ($\lambda=0$). 

Here we stress an important aspect which is evident in   Fig.~\ref{fig:virtual-zne}(a):  the statistical uncertainty associated to the PER estimates decreases for larger values of $\lambda$. For this reason PER can be considered as a low-cost error mitigation method in the sense that, compared to PEC, it requires a smaller number of samples at the price of a partial mitigation of the noise. Equivalently, for a fixed number of samples, PER results are characterized by a smaller statistical uncertainty compared to PEC.
For this reason PER could be useful in real-world  scenarios  characterized by a large level of noise, in which the sampling cost of PEC would be too large. We mention that a similar bias-variance trade off in probabilistic error cancellation has been recently studied also in Eq.~\cite{piveteau2101quasiprobability}, although with a different approach in which noise scaling is not involved.

\subsection{Virtual ZNE}
\label{sec:per-b}
A possible way to get a better estimate of the ideal expectation value is to consider PER as a noise scaling method and to apply ZNE as a second step (see also Ref.~\cite{cai2021multi}).
A demonstration of this virtual zero-noise extrapolation technique is reported in Fig.~\ref{fig:virtual-zne}(b).

The potential advantage with respect to standard ZNE is quite evident: we can now explore the region $0<\lambda < 1$ and this makes the extrapolation to $\lambda=0$ less biased.
But what is the advantage with respect to standard PEC? In PEC, the ideal expectation value  ($\lambda=0$) is directly computed with an unbiased estimator such that no extrapolation is necessary. As discussed in the case of PER, also virtual ZNE can have a smaller statistical error compared to PEC, as shown the example of Fig.~\ref{fig:virtual-zne}(b). A notable fact is that the zero-noise limit could also be extrapolated with a  non-linear function of the noisy expectation values (e.g. when using an exponential fit), while PEC is always linear by construction.

In conclusion, we may summarize both PER and virtual ZNE as hybrid techniques interpolating between two standard inference methods: ZNE (characterized by low variance but biased) and PEC (unbiased but affected by a large variance).

\subsection{Example: depolarizing-channel error reduction}
\label{subsec:depo-per}

To clarify the probabilistic error reduction (PER) method, we 
explicitly present a simple single-qubit example. We consider  the noisy gate set corresponding to all single-qubit unitaries followed by depolarizing noise: $\mathcal I = \{ \mathcal D_p \circ  \mathcal G \;|\;  \mathcal G \text{ is unitary}\}$.
Instead of completely cancelling the noise as in PEC, in PER we are interested in the effective implementation of noise scaled gates:
\begin{align} 
\mathcal G^{(\lambda)}= D_{\lambda p} \circ \mathcal G, \quad  0 \le \lambda \le \lambda_{\rm max}=3/4 p^{-1}.
\end{align}

Extending the same calculations of Refs.~\cite{temme2017error, takagi2020optimal} to the case  $\lambda\neq0$ and using the notation  $\epsilon=4/3 p$, it is easy to derive the following quasi-probability representation which is valid for any noise-scaled gate:

\begin{align} \label{eq:per_depolarizing}
 \mathcal G^{(\lambda)}  = \eta_1 \mathcal O_1 + \eta_2 \mathcal O_2 + \eta_3 \mathcal O_3 + \eta_4 \mathcal O_4,
\end{align}
where:
\begin{align} 
\eta_1 &=\left(1 + \frac{3}{4} \frac{\epsilon (1-\lambda)}{1- \epsilon} \right ), & \mathcal O_1 = \mathcal D_p \circ  \mathcal G, \nonumber \\
\eta_2 &=- \frac{1}{4}\frac{\epsilon (1-\lambda)}{1- \epsilon} , &  \mathcal O_2 = \mathcal D_p \circ    (\mathcal X \circ \mathcal G),  \nonumber \\
\eta_3 &=-\frac{1}{4}\frac{\epsilon (1-\lambda)}{1- \epsilon} , &  \mathcal O_3 = \mathcal D_p \circ    (\mathcal Y \circ \mathcal G),  \nonumber \\
\eta_4 &=- \frac{1}{4}\frac{\epsilon (1-\lambda)}{1- \epsilon} , &  \mathcal O_4  = \mathcal D_p \circ  (\mathcal Z \circ \mathcal G). \label{eq:per_etas}
\end{align}
In the definition of the noisy operations $\mathcal O_j$ given in the equation above, the composition of the unitary channels applied before $\mathcal D_p$ represent a single elementary  gate.
For $\lambda=0$, Eq.~\eqref{eq:per_etas} reduces to the PEC  quasi-probability representation obtained in Refs.~\cite{temme2017error, takagi2020optimal}. For $1\le \lambda \le \lambda_{\rm max}=\epsilon^{-1}$, all coefficients are positive corresponding to the direct application of depolarizing errors to scale up the noise (see {\it e.g.}~\cite{li2017efficient}). For $0\le \lambda \le 1$ instead the situation is less trivial and corresponds to the probabilistic error reduction regime considered in this work. In this case, the last 3 coefficients of Eq.~\eqref{eq:per_etas} are negative and the negative volume of the quasi-distribution decreases linearly with $\lambda$, interpolating between the two extremes of full error cancellation (at $\lambda=0$) and no-mitigation (at $\lambda=1$).
More explicitly the one-norm of the quasi-probability given in  
Eq.~\eqref{eq:per_etas}, as a function of $\lambda$, is given by:
\begin{equation}\label{eq:depolarizing-one-norm}
\gamma^{(\lambda)} =\Big\{
\begin{array}{lr}
\gamma - \lambda (\gamma - 1), & \lambda  \in [0, 1],  \\
1, & \lambda  \in [1, \epsilon^{-1}],
\end{array}
\end{equation}
where $\gamma=1 +\frac{3}{2} \frac{\epsilon}{1 - \epsilon}$ is the PEC one-norm (at $\lambda=0$). For the sake of completeness, we mention that Eqs.~\eqref{eq:per_etas} are also valid in the region $\lambda =[\epsilon^{-1}, p^{-1}]$ where the noise scaled depolarizing channel is completely positive (physical). However the one-norm $\gamma^{(\lambda)}$ remains equal to 1 only for  $\lambda=[\epsilon^{-1}, \frac{\gamma +1}{\gamma -1}]$, while it increases again for $\lambda \in [\frac{\gamma +1}{\gamma -1}, p^{-1}]$. This is consistent with the more general analysis that we will present in the next subsection  (Sec.~\ref{sec:canonical}, Eq.~\eqref{eq:canonical-noise-scaling-bis}).

For a fixed number of samples, Eq.~\eqref{eq:depolarizing-one-norm} implies that PER is affected by a decreasing statistical uncertainty for increasing values of $\lambda$, at the price of a larger bias error. For the same reason, the statistical variance of the virtual ZNE technique is reduced too.  Both phenomena are clearly visible in the numerical example shown in Fig.~\ref{fig:virtual-zne} and these qualitative features are reported in Table \ref{table}. 

\subsection{Canonical noise scaling}
\label{sec:canonical}

In the previous example we considered a depolarizing noise model which depends on a single parameter $p$ and therefore admits a natural notion of noise reduction $p \rightarrow \lambda p$. The same approach could be applied when the noise model is an  amplitude damping channel \cite{temme2017error, takagi2020optimal}. However, how can one meaningfully scale an arbitrary noise model? In this section we show that a generic quasi-probability representation of an ideal gate induces an abstract canonical noise model that we can associate to that gate. This abstract noise model is a well-defined  trace-preserving and completely positive channel which depends linearly on a single noise parameter $p$. In practice, this implies that noise scaling can always be meaningfully defined for any quasi-probability representation of an ideal gate.

Let us consider the representation of an ideal gate $\mathcal G$ as a linear combination of noisy operations ${\mathcal O_\alpha}$ weighted by a quasi-probability distribution $\eta_\alpha$. The explicit formula is given
in Eq.~\eqref{eq:qpr} where, for simplicity, we now drop the gate index $i$.
We now split the domain of the quasi-distribution $\eta_\alpha$ in two parts corresponding to the regions where $\eta_\alpha$ is positive and negative respectively:

\begin{equation}
D^{(\pm)} = \{\alpha, \text{ such that } \eta_\alpha \gtrless 0 \}.
\end{equation}

In a similar way we can define two completely positive and trace-preserving channels associated to the positive and negative regions respectively (see e.g. Ref.~\cite{takagi2020optimal} or, in the context of magic states,  Ref.~\cite{seddon2019quantifying}):
\begin{equation}
\Phi^{(\pm)} = \frac{1}{\gamma^{(\pm)}}\sum_{D^{(\pm)}} |\eta_\alpha| \mathcal O_\alpha, \qquad \gamma^{(\pm)} = \sum_{D^{(\pm)}} |\eta_\alpha|,
\end{equation}
where $\gamma^{(\pm)}$ are to the positive and negative volumes of the quasi-distribution, such that the 
total one-norm is $\gamma= \gamma^{(+)} + \gamma^{(-)}$, while the normalization implies that $\gamma^{(+)} - \gamma^{(-)} = 1$. 

Given the previous definitions we can reduce the multi-term linear combination in Eq.~\eqref{eq:qpr} to a linear combinations of just two channels:
\begin{equation} \label{eq:two-channels-rep}
\mathcal G = \gamma^{(+)} \Phi^{(+)} - \gamma^{(-)} \Phi^{(-)},
\end{equation}
where both channels $\Phi^{(\pm)}$ can be applied on the noisy hardware since they are a convex combination of implementable operations. What is the physical meaning of these channels? As we are going to show, $\Phi^{+}$ can be considered as a noisy approximation of $\mathcal G$, while $\Phi^{-}$ could be considered an error term that needs to be subtracted from $\Phi^{+}$ in order to recover the ideal gate $\mathcal G$.

This intuition suggests the following canonical noise model associated to the  ideal gate $\mathcal G$:
\begin{equation}\label{eq:canonical-channel}
\Lambda_p = (1 - p)\, \mathcal G + p\, \Phi^{(-)}, \quad p \in [0, 1],
\end{equation}
corresponding to a channel in which with probability $p$ the error operation $\Phi^{(-)}$ occurs, while with probability the $(1 - p)$ the ideal gate is applied without any errors.
It is easy to check that, depending the single parameter $p$, the channel $\Lambda_p$ interpolates between the three channels that appear in Eq.~\eqref{eq:two-channels-rep}:
for $p=0$ we obtain the ideal gate $\mathcal G$, for $p=\gamma^{(-)}/\gamma^{(+)}$ we obtain $\Phi^{(+)}$  which therefore could be considered as a noisy (implementable) approximation of the ideal gate, while for $p=1$ we get $\Phi^{(-)}$ which can be considered as the maximum noise limit. 

If we identify $\tilde p=\gamma^{(-)}/\gamma^{(+)}$ as the noise level of the hardware, we directly obtain a natural definition of a noise scaled gate:

\begin{equation}\label{eq:canonical-noise-scaling}
\mathcal G^{(\lambda)} = \Lambda_{\lambda \tilde p} = (1 - \lambda \tilde p)\, \mathcal G + \lambda \tilde p\, \Phi^{(-)}, \quad   \tilde p = \frac{\gamma^{(-)}}{\gamma^{(+)}},
\end{equation}
where $0 \le \lambda \le \lambda_{\rm max}= \gamma^{(+)}/\gamma{(-)}$.
By construction, this type of noise reduces the negativity of the quasi-probability representation. Indeed, replacing Eq.~\eqref{eq:two-channels-rep} into Eq.~\eqref{eq:canonical-noise-scaling}, we obtain the following representation for the
noise scaled gate in terms of implementable operations:

\begin{equation}\label{eq:canonical-noise-scaled-rep}
\mathcal G^{(\lambda)} = 
(\gamma^{(+)} - \lambda \gamma^{(-)}) \Phi^{(+)}
-(1 - \lambda) \gamma^{(-)}\Phi^{(-)},
\end{equation}
whose sampling cost (one-norm) decreases linearly with respect to the noise scale factor $\lambda$:

\begin{equation}\label{eq:canonical-noise-scaling-bis}
\gamma^{(\lambda)} =\Big\{
\begin{array}{lr}
\gamma - \lambda (\gamma - 1), & \lambda  \in [0, 1],  \\
1, & \lambda  \in [1, \lambda_{\rm max }= \frac{\gamma + 1}{\gamma - 1}].
\end{array}
\end{equation}

{\noindent \bf Note:} 
 For the optimal PEC representation of a gate in the presence of depolarizing noise given in Eq.~\eqref{eq:per_depolarizing} (evaluated at $\lambda=0$), the canonical noise channel defined in Eq.~\eqref{eq:canonical-channel} is equal to the actual physical depolarizing channel acting on the system.
In other words, the noise-scaled quasi-probability decomposition that was manually computed in
Eqs.~\eqref{eq:per_etas} is a particular case of Eq.~\eqref{eq:canonical-noise-scaled-rep}. The advantage of 
Eq.~\eqref{eq:canonical-noise-scaled-rep} is that it is well defined even for an arbitrary PEC representation, {\it e.g.}, one based on the experimental tomography of a noisy gate set.

\section{Minimum sampling cost of exact NEPEC representations}
\label{sec:gamma}
In the previous sections we presented several techniques in which ideal gates are expanded in a  noise-extended basis with {\it approximate} representations, {\it i.e.}, with a non-zero bias.
In this section instead we focus on {\it exact}  gate representations and we study the possibility of reducing the optimal sampling cost $\gamma^{\rm opt}$ introduced in Eq.~\eqref{eq:gamma_opt} by using NEPEC instead of PEC.  Our analysis is inspired by the resource theory of error mitigation proposed in Ref.~\cite{takagi2020optimal} and we follow a similar (but not identical) notation. Let $\mathcal I$ be the set of implementable operations used in PEC. Then, by noise scaling, we obtain  the extended set of implementable operations $\tilde {\mathcal I}$ introduced in Eq.~\eqref{eq:extended-implementable-ops}, such that
any ideal gate $\mathcal G_i$ of a circuit can be expanded in this extended 
basis as described in Eq.~\eqref{eq:extended-qpr}.
The associated optimal sampling cost is:
\begin{align}\label{eq:tilde_gamma_opt}
\tilde \gamma_{i}^{\rm opt} &= \min_{\substack{ \{\eta_{i,\alpha, \lambda} \} \\ \{\mathcal O_\alpha^{(\lambda)}\}}} 
\left [ \sum_{\alpha, \lambda} |\eta_{i, \alpha, \lambda}| \right] \text{ s.t. Eq. \eqref{eq:extended-qpr} holds}
\le \gamma_i^{\rm opt},
\end{align}
where the last inequality sign is a consequence of the fact that, since $\mathcal I \subset \tilde {\mathcal I}$, the minimization in Eq.~\eqref{eq:tilde_gamma_opt} is over a larger landscape w.r.t.~Eq.~\eqref{eq:gamma_opt}  and so, in principle, a smaller value can be reached. But is this inequality strict? We will show that this depends on the particular noise model and on the specific gate set $\mathcal I$.

\subsection{A no-go test}

\noindent {\bf Hypothesis 1:} \textit{Let us consider the particular setting in which each noise-scaled operation in $\tilde {\mathcal I}$ can be represented as a convex combination of operations which are in the non-scaled set $\mathcal I$, i.e.:}

\begin{equation}\label{eq:assumption-1}
\quad \forall \mathcal O^{(\lambda)} \in \tilde{\mathcal I}, \quad \mathcal O^{(\lambda)} = \sum_\alpha \mu_\lambda(\alpha) \mathcal O_\alpha, \quad  \mathcal O_\alpha \in \mathcal I,
\end{equation}
\textit{where $\{ \mu_\lambda(\alpha)\ge 0 \}$ is a positive probability distribution with respect to $\alpha$.}

In practice, if Hypothesis 1 holds, instead of actually scaling the hardware noise one could effectively obtain the same result by just probabilistically drawing an operation in $\mathcal I$ according to the probability distribution $\mu_\lambda(\alpha)$.
It is reasonable to expect that, in this case, there cannot be any advantage with respect to the sampling cost and so the inequality in \eqref{eq:tilde_gamma_opt} saturates to a trivial equality.

Indeed, by replacing Eq.~\eqref{eq:assumption-1} into Eq.~\eqref{eq:extended-qpr}, we see that for each noise-extended representation of a gate $\mathcal G_i$ there exists an equivalent representation which does not require noise scaling and has the same one-norm parameter $\gamma_i$.
The corresponding quasi-distribution is:
\begin{equation}\label{eq:product-qpr}
\eta_{i, \alpha'} = \sum_{\alpha, \lambda} \eta_{i, \alpha, \lambda}\,  \mu_{i, \alpha, \lambda}(\alpha'),
\end{equation}
whose one-norm is the same as the NEPEC one-norm, since 
\begin{equation}\sum_{\alpha'} |\eta_{i, \alpha'}|=\sum_{\alpha, \lambda} |\eta_{i, \alpha, \lambda}|\, \sum_{\alpha'} |\mu_{i, \alpha, \lambda}(\alpha')|= \sum_{\alpha, \lambda} |\eta_{i, \alpha, \lambda}|=\gamma_i.
\end{equation}
Therefore we can conclude that, if Hypothesis 1 holds, noise-scaling cannot reduce the sampling cost, i.e., $\tilde \gamma^{\rm opt}=\gamma^{\rm opt}$.\\

A relevant example in which this no-go result applies is when the set of implementable operations is $\mathcal I = \{ \mathcal E_p \circ \mathcal G, \forall \text{ unitary }  \mathcal G \} $ and $\mathcal D_p$ 
is a depolarizing channel acting on $k$ qubits (see e.g.~\cite{takagi2020optimal} for a detailed analysis of this error mitigation scenario). The corresponding noise-extended set would be  $\tilde {\mathcal I} = \{ \mathcal D_{\lambda p} \circ \mathcal G, \forall \text{ unitary }  \mathcal G, \lambda \in [1, \lambda_{\rm max} = (1- 4^{-k}) p^{-1}]\}$, where  the explicit expression for $\mathcal D_{\lambda p}$ was already given in Eq.~\eqref{eq:depo-scaled}. For $\lambda \in [1, \lambda_{\rm max}]$, $\mathcal D_{\lambda p} = \mathcal D_p \circ \mathcal D_{p'(\lambda)}$ for some $p'(\lambda) \in [0, 1]$, and so we have:
\begin{align}
\mathcal G^{(\lambda)}&=D_{\lambda p} \circ \mathcal G = \mathcal D_p \circ \mathcal D_{p'(\lambda)}\circ \mathcal G \nonumber \\
&= [1 - p'(\lambda)] \mathcal D_p \circ \mathcal G + p'(\lambda) \sum_{\mathcal P \in P_k} \frac{\mathcal D_p \circ \mathcal P \circ \mathcal G}{4^{k} - 1}, \label{eq:depo-prob-scaling}
\end{align}
where the last sum is over all the Pauli strings different from the identity as described in Eq.~\eqref{eq:depo-scaled}. Since $\mathcal P \circ \mathcal G$ is a unitary, $D_p \circ \mathcal P \circ \mathcal G \in \mathcal I$ and so Eq.~\eqref{eq:depo-prob-scaling} corresponds to a convex combination of operations which are in $\mathcal I$. This means that Hypothesis 1 holds and so, in this case, noise scaling cannot help in reducing the mitigation cost of an exact representation.\\

\subsection{Example: NEPEC representations with amplitude damping noise}
\label{subsec:amplitude-damping-example}

An example in which $\tilde \gamma^{\rm opt}<\gamma^{\rm opt}$, is given by the following noisy gate set: $\mathcal I = \{ \mathcal A_p \circ \mathcal G, \forall \text{ single-qubit unitary }  \mathcal G \} $ where $\mathcal A_p$ 
is a single-qubit amplitude damping channel. It is easy to check that, since all the $\mathcal G$ that appear in the definition of $\mathcal I$ are  unitary, 
it is impossible to obtain a valid PEC representation with this gate set \cite{temme2017error}. Indeed any linear combination of the noisy elements of  $\mathcal I$, if applied to the maximally mixed state, will always simplify to the non-unital channel  $\mathcal A_p$ and so it can never represent the action of an ideal unitary gate. For this reason, the operation ${\rm RESET}=|0 \rangle\langle 0| + |0 \rangle\langle 1|$ was used in previous works \cite{temme2017error, endo2018practical, takagi2020optimal} to make PEC feasible in the presence of amplitude damping noise.

In this proof-of-principle example   we assume that, for some reason, we cannot apply the RESET gate (e.g. because its physical implementation is too noisy or too slow).
In this case, the mitigation norm  $\gamma^{\rm opt}$ is infinite, in the sense that the minimization problem in Eq.~\eqref{eq:gamma_opt} is unfeasible without noise scaling.
On the other hand, if we extend the gate set $\mathcal I$ by noise scaling, the minimization problem in Eq.~\eqref{eq:gamma_opt} becomes feasible and we obtain a finite mitigation cost $\tilde \gamma^{\rm opt}<\gamma^{\rm opt} = \infty $.
This can be formally deduced by simply observing that $\mathcal A_{\lambda p}$, in the maximum noise limit $\lambda \rightarrow  p^{-1}$, tends to the RESET gate and so we can recover the results of Refs.~\cite{temme2017error, endo2018practical, takagi2020optimal} in this limit. Moreover we numerically find that even using intermediate values of noise scaling  $1 < \lambda <  p^{-1}$, the representation problem becomes feasible without a RESET gate at the cost of obtaining a larger (but finite) one-norm. Unfortunately we also report that, when using noise scaling on an initial gate set $\mathcal I$ which already includes the RESET gate, we could not obtain any further reduction of the sampling cost.

Inspired by the gate extrapolation technique proposed in Sec.~\ref{sec:local-zne}, we may also ask if it is possible to obtain a representation in the form of Eq.~\eqref{eq:local-zne-qpr}, {\it i.e.}, only using a linear combination of the same gate $\mathcal G_i^{(\lambda)}$ applied at different noise scale factors. 
Differently from the depolarizing super-operator $\mathcal D_p$ which is linear in $p$, the amplitude damping super-operator $\mathcal A_p$ is non-linear in $p$. This is not a problem if we aim for an approximate (biased) representation, but obtaining an exact (unbiased) representation is less trivial.  Using the change of variable $p' = 1-\sqrt{1 - p}$, $\mathcal A_p$ becomes a quadratic function of $p'$. This implies that by scaling $p'$ with 3 scale factors $\lambda \in S=[\lambda_1, \lambda_2, \lambda_3]$ we can always extrapolate to the exact zero-noise limit using the Richardson coefficients defined in Eq.~\eqref{eq:richardson-qpr}.  Minimizing the one-norm of the quasi-probability $\{ \eta_{i, \lambda}\}$ with respect to $S$ and imposing $\lambda_j \in [0, 1 / p']$, one can obtain the optimal coefficients
$S=[1, (1 + 1/p')/2, 1/p']$.
The corresponding optimized one norm is:
\begin{align}\label{eq:amp-damp-local-zne-norm}
    \gamma_i = \frac{1 +  6 p' + p'^2}{(1 - p')^2} = \frac{9 - 8 \sqrt{1 - p} - p}{1 - p}.
\end{align}
For small $p$, the equation above scales as $1 + 4p + O(p^2)$. For  comparison, we note that this is strictly larger than the sampling cost derived in Refs.~\cite{takagi2020optimal} which scales as $1 + p + O(p^2)$. This is not surprising since Eq.~\eqref{eq:amp-damp-local-zne-norm} corresponds to a very restricted noisy basis (just a single noise-scaled gate).

We conclude noticing that, while the change of variable $p \rightarrow p'$ was useful for the theoretical derivation of Eq.~\eqref{eq:amp-damp-local-zne-norm} and for proving the existence of an exact quadratic extrapolation, it is not necessary to apply it in a practical scenario. In a real use case, one can directly scale the noise level $p$ with three different noise scale factors and just solve for the coefficients $\{\eta_{i, \lambda} \,|\, \lambda \in S \}$ in Eq.~\eqref{eq:local-zne-qpr}.  The previous theoretical analysis ensures that, in the presence of amplitude damping noise,  three different scale factors are enough to obtain an exact solution of Eq.~\eqref{eq:local-zne-qpr}.\\ \\ \\

\section{Conclusions}
\label{sec:conclusions}
We propose a general error mitigation framework---NEPEC---in which a given ideal quantum circuit is represented in terms of a quasi-probability distribution over different circuits evaluated at different noise levels.
This approach generalizes existing techniques (PEC and ZNE) and, depending on the choice of the quasi-probability distribution, gives rise to different practical implementations as summarized in  Table~\ref{table}. 

A promising implementation of NEPEC is the possibility of defining approximate quasi-probability representations of individual gates (or layers) via local extrapolation and therefore without the need of performing gate set tomography. This fact can be an important practical advantage compared to standard PEC and, by construction, it bypasses noise characterization errors. For example, in the simulation reported in  Fig.~\ref{fig:noise-agnostic-pec}, we have shown the robustness of this method against noise calibration errors.

Other specific implementations of the NEPEC approach are probabilistic error reduction (PER) and virtual ZNE. In PER, noise is reduced at intermediate scale factors $\lambda \in [0, 1]$ with the advantage of requiring a smaller sampling cost compared to PEC ($\lambda=0$). Moreover, PER can also be considered as a virtual noise scaling method, and ZNE can be applied as a second post-processing step as shown in Fig.~\ref{fig:virtual-zne} (see also \cite{cai2021multi}). As a by-product of our theoretical analysis, we also identified a  canonical noise channel that can be associated to any quasi-probability representation of a gate (Sec.~\ref{sec:canonical}). This canonical channel is useful to apply PER with any noisy gate set and might be of independent interest beyond the scope of this work.

Finally we investigated if, by extending the basis of implementable operations via noise-scaling, it is possible to obtain exact gate representations with a smaller one-norm (sampling cost) compared to standard PEC representations. We found that there is a large class of situations (see Hypothesis 1 in Sec.~\ref{sec:gamma}) in which this is impossible, however, we also gave an explicit example in which NEPEC provides an advantage (Sec.~\ref{subsec:amplitude-damping-example}).

We hope that this work can stimulate further theoretical and experimental research lines. Stacking and hybridizing different quantum error mitigation techniques seems a promising strategy \cite{cai2021multi, sun2021mitigating, lowe2020unified, bultrini2021unifying}. 
The freedom of sampling over different circuits and different noise levels opens up a large variety of possible error mitigation techniques. In this work we explicitly proposed only some of them (see Table~\ref{table}), but new and perhaps better techniques  based on the NEPEC framework could be found in the future. 
Moreover, the experimental implementation of NEPEC with real quantum processors remains and open and important research task which could be addressed in the near-future.
In this regard, we comment that an experimental implementation of NEPEC is feasible with current technology. Indeed, the main technical  requirements (sampling and noise scaling) have been already achieved and experimentally demonstrated in the context of PEC \cite{zhang2020error} and ZNE \cite{kandala2019nature}.
\\

\section*{Acknowledgements}
We thank Ryan LaRose, Sarah Kaiser, Daniel Strano and Tudor Giurgica-Tiron for discussions and feedback on this work. This material is based upon work supported by the U.S. Department of Energy, Office of Science, Office of Advanced Scientific Computing Research, Accelerated Research in Quantum Computing under Award Number de-sc0020266.

\bibliographystyle{natbst}
\bibliography{references.bib}

\end{document}